\documentclass[conference]{IEEEtran}
\IEEEoverridecommandlockouts
\usepackage{cite}
\usepackage{amsmath,amssymb,amsfonts}
\usepackage{bm}
\usepackage{graphicx}
\usepackage{siunitx}
\DeclareSIUnit{\molar}{M}
\usepackage{hyperref}
\hypersetup{hidelinks}
\usepackage{tabularx}
\usepackage{titlesec}
\usepackage[font=small,labelfont=bf]{caption}
\usepackage{subcaption}
\usepackage{cleveref}
\usepackage{xfrac}

\crefname{figure}{Fig.}{Figs.}
\Crefname{figure}{Figure}{Figures}
\crefname{subfigure}{Fig.}{Figs.}
\Crefname{subfigure}{Figure}{Figures}
\crefname{table}{Table}{Tables}
\Crefname{table}{Table}{Tables}
\crefname{equation}{Eq.}{Eqs.}
\Crefname{equation}{Equation}{Equations}
\usepackage[acronym, shortcuts]{glossaries}

\newacronym{DBMC}{DBMC}{diffusion-based molecular communication}
\newacronym{TDMA}{TDMA}{time-division multiple access}
\newacronym{MDMA}{MDMA}{molecule-division multiple access}
\newacronym{NOMA}{NOMA}{non-orthogonal multiple access}
\newacronym{ADMA}{ADMA}{amplitude-division multiple access}
\newacronym{MA}{MA}{multiple access}
\newacronym{BEP}{BEP}{bit error probability}
\newacronym{TX}{TX}{transmitter}
\newacronym{RX}{RX}{receiver}
\newacronym{ISI}{ISI}{inter-symbol interference}
\newacronym{MAI}{MAI}{multiple-access interference}
\newacronym{SIC}{SIC}{successive interference cancellation}
\newacronym{OOK}{OOK}{on-off-keying}
\newacronym{MCS}{MCS}{Monte Carlo simulation}
\newacronym{SNR}{SNR}{signaling-molecule-to-noise ratio}
\newacronym{MC}{MC}{molecular communication}
\newacronym{IoBNT}{IoBNT}{internet of bio-nano-things}
\newacronym{BNM}{BNM}{bio-nano-machine}
\newacronym{DBMC-NOMA}{DBMC-NOMA}{NOMA for DBMC networks}
\newacronym{UCA}{UCA}{uniform concentration assumption}
\newacronym{SCRN}{SCRN}{stochastic chemical reaction network}
\newacronym{CRN}{CRN}{chemical reaction network}
\newacronym{ODE}{ODE}{ordinary differential equation}
\newacronym{SSA}{SSA}{stochastic simulation algorithm}
\newacronym{ChemSICal}{ChemSICal}{chemical reaction network for successive interference cancellation}
\newacronym{PDF}{PDF}{probability density function}
\newacronym{PMF}{PMF}{probability mass function}
\newacronym{RRC}{RRC}{reaction rate constant}
\newacronym{APC}{APC}{antigen-presenting cell}
\newacronym[\glsshortpluralkey={T$_\mathrm{reg}$}]{Treg}{T$_\mathrm{reg}$}{regulatory T cell}
\newacronym[longplural={effector T cells},shortplural={T$_\mathrm{eff}$}]{Teff}{T$_\mathrm{eff}$}{effector T cell}
\newacronym{IL2}{IL-2}{interleukin-2}
\newacronym{3D}{3D}{three-dimensional}

\glsdisablehyper

\def\BibTeX{{\rm B\kern-.05em{\sc i\kern-.025em b}\kern-.08em
    T\kern-.1667em\lower.7ex\hbox{E}\kern-.125emX}}

\usepackage{etoolbox}

\newcommand{\aftertitlespace}{-2.5em}

\titlespacing\section{0pt}{4pt plus 1pt minus 2pt}{3pt plus 1pt minus 2pt}
\titlespacing\subsection{0pt}{3pt plus 1pt minus 2pt}{1pt plus 1pt minus 0pt}

\AtBeginEnvironment{figure}{%
    \setlength{\abovecaptionskip}{3pt}
    \setlength{\belowcaptionskip}{0pt}
}
\AtBeginEnvironment{figure*}{%
    \setlength{\abovecaptionskip}{3pt}%
    \setlength{\belowcaptionskip}{0pt}%
}

\AtBeginEnvironment{table}{%
    \setlength{\abovecaptionskip}{0pt}
    \setlength{\belowcaptionskip}{3pt}
}
\AtBeginEnvironment{table*}{%
    \setlength{\abovecaptionskip}{0pt}%
    \setlength{\belowcaptionskip}{3pt}%
}

\renewcommand{\arraystretch}{1.03}

\usepackage[absolute]{textpos}
  
\TPMargin{5pt}

\newcommand{\copyrightstatement}{
    \begin{textblock}{0.84}(0.08,0.01)    
         \noindent
         \footnotesize
         Personal use of this material is permitted. This work has been submitted to the IEEE for possible publication. Copyright may be transferred without notice, after which this version may no longer be accessible.
    \end{textblock}
}

\begin{document}

\copyrightstatement

\setlength{\abovedisplayskip}{3pt}
\setlength{\belowdisplayskip}{3pt}

\title{An Information-Theoretic Perspective on Regulatory T Cell Localization as an IL-2 Interference Filter
\vspace{-0.5cm}
}



\author{
\IEEEauthorblockN{
Alexander Wietfeld\IEEEauthorrefmark{1},
Kristina A. Ganzinger\IEEEauthorrefmark{3},
Wolfgang Kellerer\IEEEauthorrefmark{1},
Andrew W. Eckford\IEEEauthorrefmark{2}
}
\IEEEauthorblockA{
\IEEEauthorrefmark{1}
Chair of Communication Networks, Technical University of Munich, Munich, Germany
}
\IEEEauthorblockA{
\IEEEauthorrefmark{3}
Physics of Cellular Interactions, Autonomous Matter Department, AMOLF, Amsterdam, The Netherlands}
\IEEEauthorblockA{
\IEEEauthorrefmark{2}
Department of Electrical Engineering and Computer Science, York University, Toronto, Canada
}
}

\maketitle

\vspace{\aftertitlespace}

\begin{abstract}
Diffusible signals enable communication between immune cells but can also create interference between nearby activation sites. In lymph nodes, effector T cells (T$_\mathrm{eff}$) concentrate around antigen-presenting cells (APCs), release interleukin-2 (IL-2) following antigen recognition, and sense IL-2 through cell-surface receptors. Experiments show high-uptake regulatory T cells (T$_\mathrm{reg}$) clustered around IL-2-producing cells. However, it is unclear how strongly responses at neighboring activation sites are coupled through diffusible IL-2, and how this coupling is influenced by the spatial organization of T$_\mathrm{reg}$ acting as IL-2 sinks. We interpret two neighboring sites as a molecular communication system and study the effect of clustering using a three-dimensional IL-2 field coupled to T cell population dynamics. Across a broad parameter sweep, clustering T$_\mathrm{reg}$ produces a larger mean reduction in neighboring-site interference than adding dispersed T$_\mathrm{reg}$. Compared with no T$_\mathrm{reg}$, clustered T$_\mathrm{reg}$ suppress interference while largely preserving the strong local response with corresponding consequences for population expansion and survival. Information-theoretic analysis shows improved discrimination between weak and strong local sources at short separation, whereas at larger separation clustering primarily reduces the remaining estimated neighboring-source information. These findings support a spatial-filtering role for T$_\mathrm{reg}$ clustering in maintaining locally informative immune responses.
\end{abstract}


\section{Introduction}\label{sec:introduction}

Lymph nodes bring together immune cells that detect and respond to antigens. \Acp{APC} display antigen-derived molecular fragments to T cells, which have the task of discriminating between host (self) and foreign (non-self) antigens. We call an \ac{APC} and the surrounding responding T cells an \emph{activation site}: following antigen recognition, \acp{Teff} at the activation site release \ac{IL2}, a diffusible cytokine that supports T cell division and survival, and sense it through cell-surface receptors~\cite{tkachCellsTranslateIndividual2014,voisinneCellsIntegrateLocal2015}. Neighboring activation sites therefore share a chemical medium, raising the question of how each can maintain a response that reflects its own local stimulation.

As with nearby wireless links sharing a frequency, signals from one activation site can influence reception at another. Shared \ac{IL2} can coordinate T cell responses, but also promote responses in weakly stimulated cells near a stronger source~\cite{voisinneCellsIntegrateLocal2015,tkachCellsTranslateIndividual2014}. However, \acp{Treg} strongly consume \ac{IL2} and act as local sinks~\cite{busseCompetingFeedbackLoops2010,feinermanSinglecellQuantificationIL22010}. Imaging studies have identified their clustering around \ac{IL2}-producing T cells~\cite{liuImmuneHomeostasisEnforced2015,wongLocalRegulatoryCell2021}. This motivates the hypothesis that localized uptake suppresses ``interference'' and helps neighboring activation sites maintain independent local readings. In this paper, we take a molecular-communication, and particularly an information-theoretic, approach to this localization: we interpret two neighboring sites as an information transfer system and ask whether \ac{Treg} clustering suppresses neighboring-site interference while preserving the local response. We study these effects using a \ac{3D} \ac{IL2} field coupled to T cell population dynamics, distinguishing information about local stimulation from neighboring-source influence.

Our work connects with an emerging body of literature, applying information-theoretic analysis to the T cell's discrimination task. Such analysis has been recently used to quantify discrimination performance in T cells \cite{ganti2020t,egan2022fluctuations}, distinguish classes of recognizable antigens \cite{achar2022universal,tiffeau2026different}, and to determine the role of kinetic proofreading in T cell activation \cite{hui2026mutual,kirby2023proofreading}. These works build on a rich foundational literature on the information theory of ligand-receptor systems \cite{pierobon2011noise,waltermann2011information,eckford2019channel}, of which the T cell receptor is an important example. More generally, information-theoretic quantities such as mutual information give a principled and model-independent measure of biological fitness \cite{bialek2012biophysics}, motivating their use in biological studies.

The main contribution and novelty of our paper is to apply information-theoretic analysis to the observed phenomenon of \ac{Treg} clustering.
Our key results elucidate three aspects of spatial filtering. First, clustering \ac{Treg} produces a larger mean interference reduction than adding dispersed \ac{Treg} at every tested operating point. Compared with no \ac{Treg}, clustered \ac{Treg} suppress interference with much smaller median suppression of the strong local response. Second, selective suppression extends to population expansion and persistence. Third, clustering improves local weak/strong discrimination at short separation, whereas at larger separation it mainly reduces the remaining estimated neighboring-source information. We also discuss the information-theoretic outlook on how these results might affect self/non-self discrimination.

\begin{figure*}[t]
    \centering
    \includegraphics[width=0.9\linewidth]{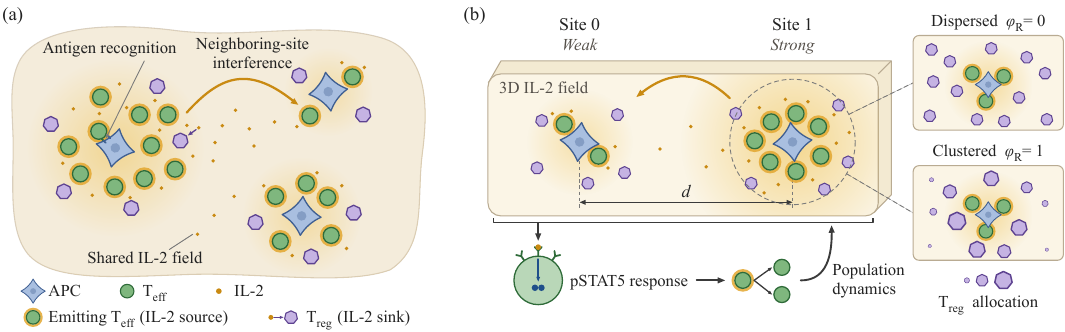}
    \caption{Biological setting and two-site model. (a) Antigen recognition at \acp{APC} initiates local \ac{Teff} responses. Shared \ac{IL2} supports these responses and causes neighboring-site interference, while \ac{Treg} act as cytokine sinks. (b) Weak and strong source sites, separated by center-to-center distance $d$, share a \ac{3D} \ac{IL2} field coupled to receptor sensing, pSTAT5 response and population dynamics. Depictions on the right compare dispersed ($\phi_{\mathrm{R}}=0$) and clustered ($\phi_{\mathrm{R}}=1$) allocation. There, \ac{Treg} symbol area represents initial local cell-equivalent mass. Cell numbers, \ac{IL2} distribution and response illustrations are schematic.}
    \label{fig:system}
\end{figure*}

\section{Physical Scenario and Communication Task}\label{sec:system_model}

Antigen recognition connects local detection to a population response, as illustrated in panel~(a) of \cref{fig:system}. \Acp{APC} present antigen-derived fragments to T cells, and recognition through the T cell receptor initiates activation~\cite{moffettComparingKineticProofreading2025}. Expansion of the responding population helps provide the cells needed for an effective immune response, while inappropriate amplification against the host's own antigens must be regulated~\cite{wongLocalRegulatoryCell2021,liuImmuneHomeostasisEnforced2015}. Our investigation concerns the cytokine-mediated amplification that follows recognition.

\subsection{Sources, Receivers, and Cytokine Sinks}

Stimulated \ac{Teff} release \ac{IL2}, which diffuses through the tissue and binds cell-surface receptors on nearby T cells. Receptor signaling activates the intracellular protein STAT5 through phosphorylation. Its phosphorylated form, pSTAT5, provides a readout of this response~\cite{feinermanSinglecellQuantificationIL22010}. Over longer timescales, \ac{IL2} supports division and survival of responding cells~\cite{tkachCellsTranslateIndividual2014,voisinneCellsIntegrateLocal2015}.

\Ac{Treg} express high-affinity \ac{IL2} receptors and compete for the cytokine~\cite{busseCompetingFeedbackLoops2010,feinermanSinglecellQuantificationIL22010}. Imaging studies identify their clustering around \ac{IL2}-producing T cells, and disrupting T cell receptor signaling in \ac{Treg} impairs this organization and control of self-reactive responses~\cite{liuImmuneHomeostasisEnforced2015,wongLocalRegulatoryCell2021}. Localized uptake can restrict signaling range~\cite{oyler-yanivTunableDiffusionConsumptionMechanism2017}, providing a mechanism through which clustering could limit neighboring-site interference.

\subsection{Communication Task in a Shared Cytokine Field}

We focus on two neighboring activation sites whose T cells share an \ac{IL2} field. Cytokine from a strong source can amplify the response at a weakly stimulated neighboring site. We ask whether \ac{Treg} clustering suppresses this neighboring-site interference while preserving a strong local response. The communication objectives are to retain information about local stimulation and reduce unwanted dependence on neighboring stimulation. These objectives concern how the responding population is amplified, and connect spatial cytokine regulation to locally appropriate immune responses.

\section{Model and Evaluation Methods}\label{sec:simulation_setup}

\subsection{Spatial Representation and Source Recruitment}

We represent two activation sites, indexed by $i\in\{0,1\}$ and separated by distance $d$, using fixed spatial supports in a \ac{3D} tissue volume (\cref{fig:system}(b)). Supports belong to their nearest site and carry cell-equivalent masses that change through division and loss. Descendants remain at their original support. Potential \ac{IL2} source positions favor proximity to the site center, with weight $\phi_{\mathrm{E}}$ relative to random placement.

Recruitment denotes the initiation of a \ac{Teff} response, including finite \ac{IL2} emission and the ability to divide. Each original recruited cell is a {\em founder}, and together with its descendants forms a {\em lineage}. Only living original founders emit. The input $X_i\in\{W,S\}$ sets the cumulative number of recruited founders to $N_i=N_{\mathrm{weak}}$ or $N_{\mathrm{on}}$, respectively, representing weak and strong source-strength conditions~\cite{fuhrmannAdequateImmuneResponse2016,tkachCellsTranslateIndividual2014,voisinneCellsIntegrateLocal2015}. Recruitment times $\tau_k$ are uniform over $[0,T_{\mathrm{rec}}]$. The four two-site inputs are $(W,W)$, $(W,S)$, $(S,W)$, and $(S,S)$.

\subsection{IL-2 Emission, Transport, and Signaling}

Founder $k$ emits at rate $q_{\mathrm{emit}}$ per living original cell-equivalent for duration $T_{\mathrm{emit}}$~\cite{fuhrmannAdequateImmuneResponse2016,wongMesoscaleCellAntigen2021}. Its flux and the distributed field source are
\begin{equation}
\begin{aligned}
F_k(t)&=q_{\mathrm{emit}}A_k^{\mathrm{o}}(t)\mathbf{1}\{0\leq t-\tau_k<T_{\mathrm{emit}}\},\\
Q(\boldsymbol{x},t)&=\sum_kF_k(t)G_{\sigma}(\boldsymbol{x}-\boldsymbol{x}_k).
\end{aligned}
\label{eq:source_deposition}
\end{equation}
Here, $A_k^{\mathrm{o}}$ is the living original-founder mass and $G_{\sigma}$ a normalized deposition kernel of width $\sigma=\SI{8}{\micro\metre}$. 
Daughter cells consume and respond to \ac{IL2} but do not emit.

We assume that the cytokine field relaxes faster than the cell populations change and use a quasi-stationary approximation~\cite{brunnerDiffusionlimitedCytokineSignaling2024}:
\begin{equation}
Q=-\nabla\!\cdot\!\left(aD\nabla c\right)+f\eta c+\sum_r\frac{V_{\max,r}c}{K_r+c}.
\label{eq:il2_field}
\end{equation}
The concentration $c(\boldsymbol{x},t)$ balances emission, diffusion $D$, decay $\eta$, and saturating uptake. Homogenization replaces resolved cell surfaces by distributed uptake, with diffusion correction $a$ and extracellular volume fraction $f$ set by cell geometry~\cite{liHomogenizationApproachSpatial2026}. The reference cell radius is \SI{5}{\micro\metre} and support spacing $\ell=\SI{12}{\micro\metre}$. Boundary exchange permits loss to the surrounding medium.

The uptake classes $r$ are unrecruited \ac{Teff} ($\mathrm{U}$), recruited lineages ($\mathrm{L}$), and baseline- and high-uptake \ac{Treg} ($\mathrm{Rb},\mathrm{Rh}$). Their capacities $V_{\max,r}$ combine internalization rate $k_{\mathrm{int}}$, effective receptor count $R_r$, living mass, and receptor variability, distributed with the source deposition width~\cite{oyler-yanivTunableDiffusionConsumptionMechanism2017,thurleyThreeDimensionalGradientsCytokine2015,brunnerDiffusionlimitedCytokineSignaling2024}. $K_r$ is the uptake half-saturation concentration. The local modeled pSTAT5 response is
\begin{equation}
p_r(c)=\frac{c^h}{\mathrm{EC}_{50,r}^{h}+c^h}.
\label{eq:pstat_response}
\end{equation}
The signaling half-saturation is $\mathrm{EC}_{50,r}$, and $h$ controls response steepness~\cite{feinermanSinglecellQuantificationIL22010}. 

\subsection{\texorpdfstring{$T_{\mathrm{reg}}$}{Treg} Localization and Population Dynamics}

A fraction $f_{\mathrm{R}}$ of supports is reserved for \ac{Treg}. Let $\mathcal{R}_i$ contain the $M_i$ reserved supports at site $i$. Define $h_i(\boldsymbol{x})$ as the sum of Gaussian weights centered at that site's fixed source positions, each with peak value 1 and standard deviation $\sigma_{\mathrm{R}}$. The initial \ac{Treg} allocation is
\begin{equation}
n_{\mathrm{R},k}(0)=(1-\phi_{\mathrm{R}})+\phi_{\mathrm{R}}\frac{M_i h_i(\boldsymbol{x}_k)}{\sum_{\ell\in\mathcal{R}_i}h_i(\boldsymbol{x}_\ell)}.
\label{eq:treg_allocation}
\end{equation}
Thus, $\phi_{\mathrm{R}}=0$ disperses one cell-equivalent per reserved support, whereas $\phi_{\mathrm{R}}=1$ concentrates the same initial total near potential sources. The profile is shared across matched source inputs. Subsequent population changes need not conserve this total. The no-\ac{Treg} control retains the reserved positions but removes their uptake.

Between recruitment events, recruited \ac{Teff} lineage mass changes through division and loss. Cytokine signaling promotes a division-capable state. Loss combines basal turnover with increased cytokine-deprivation sensitivity after division~\cite{deboerEstimatingLymphocyteDivision2006,pandiyanCD4+CD25+Foxp3RegulatoryCells2007}. Original founders undergo loss, while new mass enters non-emitting descendants. Delayed cycling and deprivation responses allow populations to contract more slowly than signaling declines.

\Ac{Treg} combine antigen-associated activation with cytokine-dependent cycling, adaptation between uptake states, and homeostatic replacement toward their initial local reservoir~\cite{busseCompetingFeedbackLoops2010,liuImmuneHomeostasisEnforced2015,wongLocalRegulatoryCell2021}. The localization parameter also biases the spatial activation program toward the source-density profile.

\subsection{Simulation Procedure and Evaluated Conditions}

At each biological step $\Delta t=\SI{1}{\hour}$, a \ac{3D} finite-volume solver recomputes the quasi-stationary field. Responses are recorded before updating cycling, division, and loss, completing one step of the feedback loop illustrated in panel~(b) of \cref{fig:system}. Simulations extend to \SI{240}{\hour}, with $0$--$192$ h used for the primary evaluation and the final $48$ h for validation.

The central operating point is $(N_{\mathrm{on}},N_{\mathrm{weak}},d)=(360,3,\SI{100}{\micro\metre})$. The broad sweep combines three strong and four weak founder counts with $d\in\{60,100,150\}\,\si{\micro\metre}$, giving 36 points. Targeted analyses use $d\in\{75,100,125\}\,\si{\micro\metre}$ at the central counts. We compare no \ac{Treg} with $\phi_{\mathrm{R}}\in\{0,0.5,1\}$.

Seeds specify position jitter (standard deviation $\sigma_x$), mean-one lognormal receptor variation (coefficient of variation $\mathrm{CV}_{\mathrm{R}}$), and recruitment assignments and times, matched across regulatory conditions~\cite{brunnerDiffusionlimitedCytokineSignaling2024}. We use 50 independent seeds per operating point. \Cref{tab:model_parameters} gives the principal parameters and evaluated conditions. The source strengths and recruitment schedule are effective model choices.

\begin{table*}[t]
\caption{Principal model parameters and evaluation settings. Sets denote evaluated scenarios, with central operating-point values underlined. \\ L: literature-derived. EL: effective model choice with literature evidence. M: model choice.}
\label{tab:model_parameters}
\centering
\footnotesize
\sisetup{per-mode=symbol}
\setlength{\tabcolsep}{4pt}
\renewcommand{\arraystretch}{1.03}
\begin{tabularx}{\textwidth}{>{\raggedright\arraybackslash}p{0.19\textwidth} c >{\raggedright\arraybackslash}X l}
\hline
Category & Symbol & Value / evaluated range & Basis \\
\hline
Site separation & $d$ & $d\in\{60,75,\underline{100},125,150\}\,\si{\micro\metre}$ & EL \cite{liuImmuneHomeostasisEnforced2015,oyler-yanivTunableDiffusionConsumptionMechanism2017,gongPredictingLymphNode2013}\\
Recruited founders & $N_{\mathrm{on}},N_{\mathrm{weak}}$ & $N_{\mathrm{on}}\in\{96,\underline{360},768\}$, $N_{\mathrm{weak}}\in\{1,\underline{3},9,27\}$ & EL \cite{quielAntigenstimulatedCD4Tcell2011,tkachCellsTranslateIndividual2014,fuhrmannAdequateImmuneResponse2016}\\
Recruitment/emission & $T_{\mathrm{rec}},T_{\mathrm{emit}},q_{\mathrm{emit}}$ & \SI{72}{\hour}, \SI{12}{\hour}, $5\,\mathrm{molecules\,s^{-1}}$ & EL \cite{fuhrmannAdequateImmuneResponse2016,wongMesoscaleCellAntigen2021}\\
Source placement & $\phi_{\mathrm{E}}$ & $0.9$ & M\\
\ac{Treg} localization & $f_{\mathrm{R}},\sigma_{\mathrm{R}}$ & $0.05$, \SI{24}{\micro\metre} & EL \cite{liuImmuneHomeostasisEnforced2015,wongLocalRegulatoryCell2021,brunnerDiffusionlimitedCytokineSignaling2024}\\
IL-2 transport & $D,\eta$ & \SI[per-mode=symbol]{10}{\micro\metre\squared\per\second}, $2.8\times10^{-5}\,\mathrm{s^{-1}}$ & L \cite{brunnerDiffusionlimitedCytokineSignaling2024}\\
Uptake capacity & $R_{\mathrm{U}},R_{\mathrm{L}},R_{\mathrm{Rb}},R_{\mathrm{Rh}}$ & $200,300,300,750$ effective receptors per cell-equivalent & EL \cite{oyler-yanivTunableDiffusionConsumptionMechanism2017,thurleyThreeDimensionalGradientsCytokine2015,brunnerDiffusionlimitedCytokineSignaling2024}\\
Receptor internalization & $k_{\mathrm{int}}$ & $4.6\times10^{-4}\,\mathrm{s^{-1}}$ & EL \cite{thurleyThreeDimensionalGradientsCytokine2015,brunnerDiffusionlimitedCytokineSignaling2024}\\
Uptake half-saturation & $K_{\mathrm{U}},K_{\mathrm{L}},K_{\mathrm{Rb}},K_{\mathrm{Rh}}$ & \SIlist{50;20;10;3}{\pico\molar} & EL \cite{busseCompetingFeedbackLoops2010,hoferCompetitionIL2Regulatory2012}\\
IL-2 response & $\mathrm{EC}_{50,\mathrm{L}},\mathrm{EC}_{50,\mathrm{R}},h$ & \SIlist{10;2}{\pico\molar}, $h=3$ & L \cite{feinermanSinglecellQuantificationIL22010,brunnerDiffusionlimitedCytokineSignaling2024}\\
\hline
Variability & $\sigma_x/\ell,\mathrm{CV}_{\mathrm{R}}$ & $0.05,0.20$ & M\\
Bootstrap replicates & --- & $2{,}000$ suppression, $1{,}000$ temporal/information & M\\
Self-prior construction & $p_P,p_A,\rho_P,\rho_A$ & Presence / activation: $0.8,0.1$, across-site correlations: $0.1,0.1$ & M\\
\hline
\end{tabularx}
\end{table*}
\subsection{Channel Outputs}\label{sec:channel_outputs}

Let $L_i(t)$ be the complete living \ac{Teff} lineage mass at site $i$, including recruited founders and descendants, and $\overline{p}_i(t)$ its mass-weighted mean modeled pSTAT5 response. The count $N_i$ includes all distinct recruited founders, including those that subsequently die. Unrecruited cells consume cytokine but are excluded from these outputs.

The signaling response averages over hourly snapshots during recruitment:
\begin{equation}
Y_i^{\mathrm{sig}}=\sum_{t=0}^{72}L_i(t)\,\overline{p}_i(t)\Big/\sum_{t=0}^{72}L_i(t).
\label{eq:licensed_pstat}
\end{equation}
This dimensionless value in $[0,1]$ gives the average signaling intensity across the recruited population. Peak expansion is the largest living lineage mass per recruited founder:
\begin{equation}
Y_i^{\mathrm{peak}}=\max_{0\leq t\leq192}L_i(t)\bigg/N_i.
\label{eq:peak_lineage}
\end{equation}
Its unit is cell-equivalents per founder. Post-recruitment burden integrates the living population over $72$--$192$ h:
\begin{equation}
Y_i^{\mathrm{burden}}=\frac{1}{N_i}\int_{72}^{192}L_i(t)\,\mathrm{d}t.
\label{eq:lineage_burden}
\end{equation}
We evaluate the integral by the trapezoidal rule on hourly samples. It measures cumulative population persistence in cell-hours per founder. Founder normalization follows established experimental normalized expansion measures~\cite{quielAntigenstimulatedCD4Tcell2011}.

\subsection{Neighboring-Site Interference and \texorpdfstring{$T_{\mathrm{reg}}$}{Treg} Filtering}\label{sec:interference_metrics}

Write $Y_{i,\phi}^{\alpha}(x_i|x_j)$ for output $\alpha\in\{\mathrm{sig},\mathrm{peak},\mathrm{burden}\}$, with local input first and neighbor input second. We use $\phi=\phi_{\mathrm{R}}$ when \ac{Treg} are present and $\phi=\varnothing$ otherwise. Neighboring-site interference is
\begin{equation}
C_{\phi}^{\alpha}=Y_{i,\phi}^{\alpha}(W\mid S)-Y_{i,\phi}^{\alpha}(W\mid W),
\label{eq:neighbor_interference}
\end{equation}
the additional weak-site response caused by a strong neighbor. The reductions from adding dispersed \ac{Treg} and subsequently clustering them are
\begin{equation}
F_{\mathrm{presence}}^{\alpha}=C_{\varnothing}^{\alpha}-C_{0}^{\alpha},\quad F_{\mathrm{clustering}}^{\alpha}=C_{0}^{\alpha}-C_{1}^{\alpha}.
\label{eq:presence_clustering}
\end{equation}
Positive values mean less interference. Relative to no \ac{Treg}, interference and strong local-response suppression are
\begin{equation}
S_{\mathrm{C}}^{\alpha}(\phi)=1-\frac{C_{\phi}^{\alpha}}{C_{\varnothing}^{\alpha}},\quad S_{\mathrm{L}}^{\alpha}(\phi)=1-\frac{Y_{i,\phi}^{\alpha}(S\mid W)}{Y_{i,\varnothing}^{\alpha}(S\mid W)}.
\label{eq:coupling_suppression}
\end{equation}

\subsection{Information About Local and Neighboring Source Strength}\label{sec:information_metrics}

We use the signaling response directly, $\widetilde{Y}_i=Y_i^{\mathrm{sig}}$, and compress the wider population-output range using $\widetilde{Y}_i=\log(1+Y_i^{\alpha})$ for $\alpha\in\{\mathrm{peak},\mathrm{burden}\}$. This monotone transformation preserves ordering and changes the classifier's fitting scale. The two information tasks are
\begin{equation}
\begin{aligned}
I_{\mathrm{local}}&=I(X_i;\widetilde{Y}_i\mid X_j=S),\\
I_{\mathrm{cross}}&=I(X_j;\widetilde{Y}_i\mid X_i=W).
\end{aligned}
\label{eq:information_tasks}
\end{equation}
The first compares $W|S$ with $S|S$ to assess local weak/strong discrimination. The second compares $W|W$ with $W|S$ to assess neighboring-source dependence. Each uses an equal binary input prior.

For continuous responses and limited seed counts, logistic regression estimates input probabilities without response binning. We partition seeds into five groups, fitting on four and evaluating on the fifth. Both site orientations are included, with all responses from one seed kept together. For either task,
\begin{equation}
\widehat{I}_{\mathrm{est}}(X;\widetilde{Y}_i)
\!=\!H(X)+\frac{1}{n}\!\!\sum_{m=1}^{n}\log_2\widehat{P}(X\!=\!x_m|\widetilde{Y}_i\!=\!y_m),
\label{eq:classifier_information}
\end{equation}
where $H(X)=1$ bit and $n$ counts held-out responses. Scoring probabilities assigned to the correct input estimates a classifier-based mutual-information lower bound. We obtain 95\% intervals by resampling seed groups, using held-out prediction losses without refitting.

\begin{figure*}[!t]
    \centering
    \includegraphics[width=0.9\textwidth]{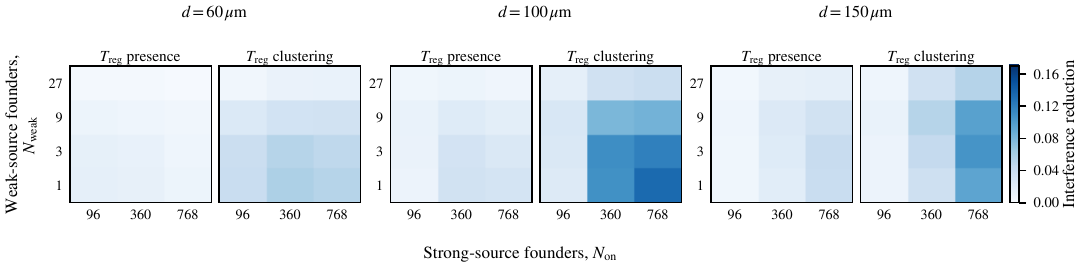}
    \caption{Interference reduction in the signaling response $Y_i^{\mathrm{sig}}$ over $0$--$72$ h. At each distance, the left map shows adding dispersed \ac{Treg} and the right map shows clustering them, as defined in \cref{eq:presence_clustering}.}
    \label{fig:filtering_map}
\end{figure*}
\subsection{Prior-Dependent Self/Non-Self Outlook}\label{sec:self_nonself_methods}

This analysis links a hypothetical self/non-self class $Z$, source pattern $\boldsymbol{X}=(X_0,X_1)$, and decoded response $Y$ through $Z\!\rightarrow\!\boldsymbol{X}\!\rightarrow\!Y$. We consider cases with at least one strong source, fixing $P(Z=\mathrm{nonself})=0.05$. In the order $(W,S),(S,W),(S,S)$, the self prior $\boldsymbol{\pi}_{\mathrm{self}}\simeq(0.4578,0.4578,0.0845)$ represents frequent but rarely activating inputs. It follows from independent presence and conditional-activation layers specified in \cref{tab:model_parameters}. For non-self, $q$ is the probability of two strong sources among triggered cases, giving $\boldsymbol{\pi}_{\mathrm{nonself}}(q)=((1-q)/2,(1-q)/2,q)$ for $q\in\{0.25,0.5,0.75\}$.

A quadratic-discriminant classifier uses the two-site peak-expansion vector $\boldsymbol{U}=[\log(1+Y_0^{\mathrm{peak}}),\log(1+Y_1^{\mathrm{peak}})]$ with seed-grouped five-fold evaluation. For each regulatory condition, averaging held-out probabilities for each true pattern gives the four-output decoder channel $P(Y|\boldsymbol{X})$. We vary the class-dependent pattern priors without changing this channel.

The data-processing relation identifies the class information lost after decoding, and we report the retained fraction:
\begin{equation}
I(Z;\boldsymbol{X})\!-\!I(Z;Y)\!=\!I(Z;\boldsymbol{X}|Y)\!\geq\!0,\ I_{\mathrm{avail}}\!=\!\frac{I(Z;Y)}{I(Z;\boldsymbol{X})}.
\label{eq:available_information}
\end{equation}
The ratio normalizes by each prior's input-information budget. 

\section{Results}\label{sec:results}

\subsection{Clustering Adds to the Filtering Effect of \texorpdfstring{$T_{\mathrm{reg}}$}{Treg} Presence}\label{sec:results_filtering_map}

We compare no \ac{Treg}, dispersed \ac{Treg}, and clustered \ac{Treg} to distinguish adding a regulatory sink from changing its spatial organization. The local response is measured at each activation site. Weak and strong sources correspond to smaller and larger numbers of recruited \ac{IL2}-producing cells. Interference is the extra response at a weak site caused by replacing its weak neighbor with a strong one.

\Cref{fig:filtering_map} evaluates this interference using the signaling response, the population-weighted mean modeled pSTAT5 intensity during $0$--$72$ h. It separates the reduction from adding dispersed \ac{Treg}, $F_{\mathrm{presence}}^{\mathrm{sig}}$, from that obtained by clustering them, $F_{\mathrm{clustering}}^{\mathrm{sig}}$. Clustering produced a larger mean reduction at all 36 tested operating points.

Compared with no \ac{Treg}, fully clustered \ac{Treg} reduced signaling interference at every operating point. Median percentage suppression across the operating-point summaries was $33.6\%$, compared with $3.75\%$ suppression of the strong local response. Thus, the combined effect of presence and clustering preferentially suppressed neighboring-site interference.

\begin{figure}[!t]
    \centering
    \includegraphics[width=\linewidth]{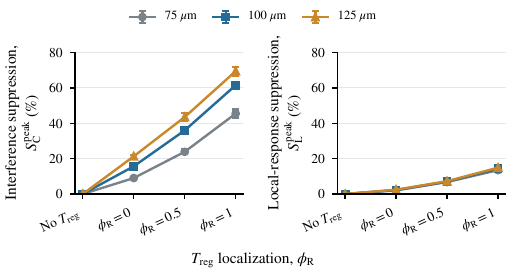}
    \caption{Suppression of interference (left) and strong local response (right) in peak expansion $Y_i^{\mathrm{peak}}$ at $(N_{\mathrm{on}},N_{\mathrm{weak}})=(360,3)$, relative to no \ac{Treg}. Points are mean suppression percentages, with 95\% bootstrap intervals.}
    \label{fig:suppression_comparison}
\end{figure}

\subsection{Selective Suppression Extends to Population Responses}\label{sec:results_multioutput}

Peak expansion is the largest living population of recruited cells and their descendants over $0$--$192$ h, divided by the number of original recruited cells, or founders. \Cref{fig:suppression_comparison} compares suppression of interference with suppression of the strong local response, both relative to no \ac{Treg}. At \SI{100}{\micro\metre}, full clustering suppressed peak-expansion interference by $61.5\%$, compared with $14.3\%$ suppression of the strong local response. Supporting analyses showed the same selective suppression for signaling and post-recruitment burden, the cumulative living population per founder over $72$--$192$ h, across separations of $75$, $100$, and \SI{125}{\micro\metre}.

\begin{figure}[!t]
    \centering
    \includegraphics[width=\linewidth]{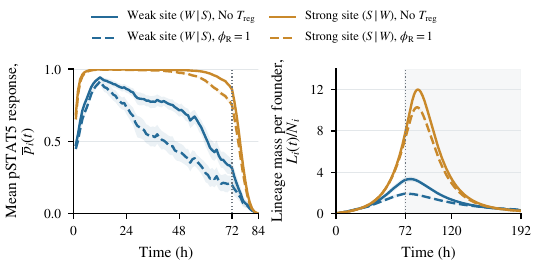}
    \caption{Mean pSTAT5 response $\overline{p}_i(t)$ (left) and lineage mass per founder $L_i(t)/N_i$ (right) at $(N_{\mathrm{on}},N_{\mathrm{weak}},d)=(360,3,\SI{100}{\micro\metre})$. Bands are 95\% bootstrap intervals. The vertical line marks the end of recruitment.}
    \label{fig:time_courses}
\end{figure}

The time trajectories connect signaling suppression to population development. In $x_i|x_j$, the local source is listed first and its neighbor second. \Cref{fig:time_courses} compares a weak site with a strong neighbor, $W|S$, and a strong site with a weak neighbor, $S|W$, at $(N_{\mathrm{on}},N_{\mathrm{weak}},d)=(360,3,\SI{100}{\micro\metre})$. Relative to no \ac{Treg}, full clustering suppresses the weak site's signaling and subsequent expansion more strongly. Recruitment ends at \SI{72}{\hour}, and emission ends by \SI{84}{\hour}. Signaling falls faster than populations contract.

\begin{figure}[!t]
    \centering
    \begin{subfigure}[t]{\columnwidth}
        \centering
        \includegraphics[width=0.9\linewidth]{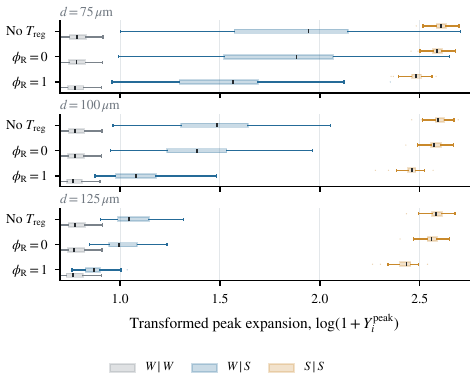}
        \caption{Transformed peak-expansion distributions. Boxes show medians and interquartile ranges, with whiskers within $1.5$ interquartile ranges.}
        \label{fig:conditional_information_distributions}
    \end{subfigure}
    \begin{subfigure}[t]{\columnwidth}
        \centering
        \includegraphics[width=\linewidth]{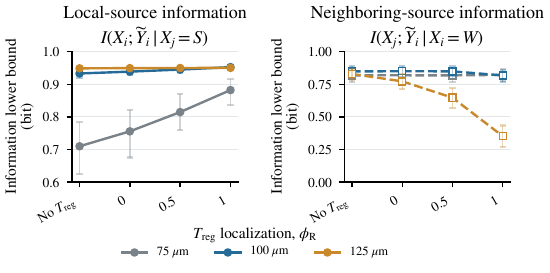}
        \caption{Local- and neighboring-source information lower bounds. Information bars are 95\% bootstrap intervals. The two information axes have different vertical ranges.}
        \label{fig:conditional_information_bounds}
    \end{subfigure}
    \caption{Distance-dependent information effects for peak expansion at $(N_{\mathrm{on}},N_{\mathrm{weak}})=(360,3)$. }
    \label{fig:conditional_information}
    \vspace{-0.2cm}
\end{figure}

\subsection{Distance Changes the Dominant Information Benefit}\label{sec:results_information}

\Cref{fig:conditional_information_distributions} places three peak-expansion distributions on a logarithmic response axis. The comparison $W|S$ versus $S|S$ asks whether a neighbor-supported weak response can still be distinguished from a genuinely strong local response. The comparison $W|W$ versus $W|S$ asks how strongly the weak site's response still depends on its neighbor. Higher local-source information supports the first objective, while lower neighboring-source information supports the second. At \SI{75}{\micro\metre}, full clustering suppresses the neighbor-supported weak response more than the strong local response, improving their separation.

\Cref{fig:conditional_information_bounds} quantifies these comparisons using classifier-based information lower bounds, with equally likely weak and strong inputs in each comparison. Relative to no \ac{Treg}, full clustering increased local-source information from $0.710$ to $0.882$ bit at \SI{75}{\micro\metre}, while neighboring-source information remained approximately $0.82$ bit. The weak response moved away from the strong response but remained distinguishable from the weak-neighbor baseline. At \SI{100}{\micro\metre}, estimated local-source information increases slightly while neighboring-source information begins to decline.

At \SI{125}{\micro\metre}, local-source information was already approximately $0.95$ bit. Full clustering instead reduced estimated neighboring-source information from $0.829$ to $0.352$ bit as $W|S$ approached $W|W$. The dominant benefit therefore shifts from improved local weak/strong discrimination at short separation to reduced estimated neighboring-source information at larger separation.

Adding dispersed \ac{Treg} and then clustering them increased estimated local-source information at \SI{75}{\micro\metre} by $0.046$ and $0.127$ bit, respectively, and reduced neighboring-source information at \SI{125}{\micro\metre} by $0.056$ and $0.421$ bit. Spatial organization therefore contributed more than adding a dispersed sink.

\begin{figure}[!t]
    \centering
    \includegraphics[width=\linewidth]{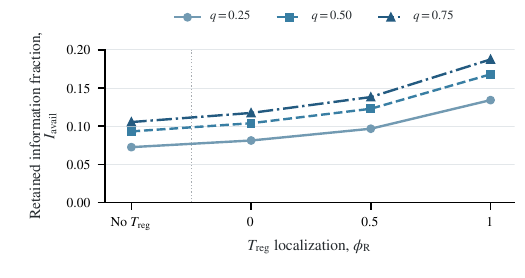}
    \caption{Retained self/non-self information fraction $I_{\mathrm{avail}}$ from two-site peak expansion at $(N_{\mathrm{on}},N_{\mathrm{weak}},d)=(360,3,\SI{60}{\micro\metre})$. Curves vary the probability $q$ of two strong sources among triggered non-self inputs, with other prior settings defined in \cref{sec:self_nonself_methods}.}
    \label{fig:declared_class_outlook}
\end{figure}

\subsection{A Prior-Dependent Self/Non-Self Outlook}\label{sec:results_self_nonself}

\Cref{fig:declared_class_outlook} returns to self/non-self discrimination by assigning hypothetical class-dependent frequencies to the same simulated source-strength patterns at \SI{60}{\micro\metre}. We consider cases with at least one strong source and fix the non-self frequency at $0.05$. The parameter $q$ varies how often both sources are strong among non-self cases, with self-pattern frequencies held fixed. The fraction $I_{\mathrm{avail}}$ in \cref{eq:available_information} measures the class information retained after decoding both sites' peak expansion, relative to that available in the source pattern.

Across the three displayed priors, retention ranged from $0.073$--$0.106$ without \ac{Treg} to $0.134$--$0.188$ under full clustering. At $q=0.5$, clustering increased retention from $0.104$ with dispersed \ac{Treg} to $0.168$. Retention increased with both clustering and $q$ over the displayed values. This outlook suggests that limiting neighboring-site interference can help preserve evidence of genuinely strong sources at both sites under assumed self/non-self input statistics.

\section{Conclusions and Outlook}\label{sec:conclusion}

This work has shown how \ac{Treg} clustering can preserve locally informative responses in a shared \ac{IL2} field. Across the tested sweep, clustering removed more signaling interference than adding dispersed \ac{Treg}. Compared with no \ac{Treg}, clustered uptake suppressed interference more strongly than the strong local response, extending to population expansion and persistence. The information benefit shifts from improved local weak/strong discrimination at short separation to reduced estimated neighboring-source information at larger separation.
The model uses effective source strengths and recruitment schedules, fixed positions, and a modeled pSTAT5 readout. Quantitative predictions will require experimental calibration. Information estimates depend on the choice of output response, prior, and decoder. Calibrated \ac{IL2}, pSTAT5, and growth measurements could test how local \ac{Treg} organization shapes neighboring responses in practice. Dynamic microdomain formation and experimentally supported priors are planned to strengthen the immune-decision 
analysis and encourage research into 
engineered cytokine sources or sinks to support local responses 
and limit
neighboring activation.

\bibliographystyle{ieeetr}
\bibliography{references}

\end{document}